\documentclass[aps,pre,twocolumn,superscriptaddress,longbibliography,notitlepage]{revtex4-1}

\usepackage{graphicx}
\usepackage{enumitem}
\usepackage{color} 
\usepackage{hyperref}
\usepackage{xcolor}
\usepackage{mathtools,amsmath,amssymb}
\usepackage{comment}

\newcommand{\bs}{\boldsymbol}

\begin{document}

\title{Emergence and melting of active vortex crystals}

\author{Martin James}
\email{martin.james@ds.mpg.de}
\affiliation{Max Planck Institute for Dynamics and Self-Organization (MPI DS), Am Fa{\ss}berg 17, 37077 G\"{o}ttingen, Germany}
\author{Dominik Anton Suchla}
\affiliation{Max Planck Institute for Dynamics and Self-Organization (MPI DS), Am Fa{\ss}berg 17, 37077 G\"{o}ttingen, Germany}
\affiliation{Faculty of Physics, University of G\"ottingen, Friedrich-Hund-Platz 1, 37077 G\"{o}ttingen, Germany}
\author{J\"orn Dunkel}
\affiliation{Department of Mathematics, Massachusetts Institute of Technology, 77 Massachusetts Avenue, Cambridge, Massachusetts 02139-4307, USA}
\author{Michael Wilczek}
\email{michael.wilczek@ds.mpg.de}
\altaffiliation{\\MJ and DAS are joint first authors}
\affiliation{Max Planck Institute for Dynamics and Self-Organization (MPI DS), Am Fa{\ss}berg 17, 37077 G\"{o}ttingen, Germany}
\affiliation{Faculty of Physics, University of G\"ottingen, Friedrich-Hund-Platz 1, 37077 G\"{o}ttingen, Germany}

\begin{abstract}

Melting of two-dimensional (2D) equilibrium crystals, from superconducting vortex lattices to colloidal structures, is a complex phenomenon characterized by the sequential loss of positional and orientational order. Whereas melting processes in passive systems are typically triggered by external heat injection, active matter crystals can self-assemble and melt into an active fluid by virtue of their intrinsic motility and inherent non-equilibrium stresses. Emergent crystal-like order has been observed in recent experiments on suspensions of swimming sperm cells, fast-moving bacteria, Janus colloids, and in embryonic tissues. Yet, despite recent progress in the theoretical description of such systems, the non-equilibrium physics of active crystallization and melting processes is not well understood. Here, we establish the emergence and investigate the melting of self-organized vortex crystals in 2D active fluids using an experimentally validated generalized Toner-Tu theory. Performing hydrodynamic simulations at an unprecedented scale, we identify two distinctly different melting scenarios: a hysteretic discontinuous phase transition and melting through an intermediary hexatic phase, both of which can be controlled by self-propulsion and active stresses. Our analysis further reveals intriguing transient features of active vortex crystals including meta-stable superstructures of opposite spin polarity. Generally, these results highlight the differences and similarities between crystalline phases in active fluids and their equilibrium counterparts.
\end{abstract}

\maketitle

Melting of crystal structures is a ubiquitous phase transition phenomenon that typically requires the injection of heat from an external source~\cite{strandburg1988two}. An intriguing exception to this rule are active non-equilibrium crystals that can both self-assemble~\cite{palacci2013living,mognetti2013living} and melt into an active fluid~\cite{klamser2018thermodynamic,weber2014defect,paliwal2020role}, owing to the intrinsic motility of their microscopic constituents. The spontaneous emergence and destruction of crystal-like order can be observed in a wide variety of natural and artificial systems~\cite{palacci2013living,petroff2015fast,DUTTA2019743,riedel2005self}. Striking examples range from suspensions of active colloids~\cite{palacci2013living,theurkauff2012dynamic,takatori2016acoustic}, bacteria~\cite{petroff2015fast} and sperm cells~\cite{riedel2005self} to biological tissues~\cite{DUTTA2019743} that can solidify and fluidize during embryonic development~\cite{Mongera:2018aa,Bi:2015aa}. Yet, despite recent experimental advances~\cite{petroff2015fast,volfson2008biomechanical,riedel2005self,DUTTA2019743,Mongera:2018aa,ramananarivo2019activity}  and important theoretical progress~\cite{paliwal2020role,praetorius2018active,singh2016universal,
van2016spatiotemporal,nguyen2014emergent,engel2013hard,menzel2014active,
menzel2013traveling,bialke2012crystallization,Mongera:2018aa,Bi:2015aa,durand2019thermally,loewe2019solid}, many key aspects of active melting processes remain poorly understood. This may not come as a big surprise given that it took several decades to decipher the complex melting scenarios of even the most basic 2D equilibrium crystal structures~\cite{strandburg1988two}.

\begin{figure*}
\centering
\includegraphics[width=1.0\textwidth]{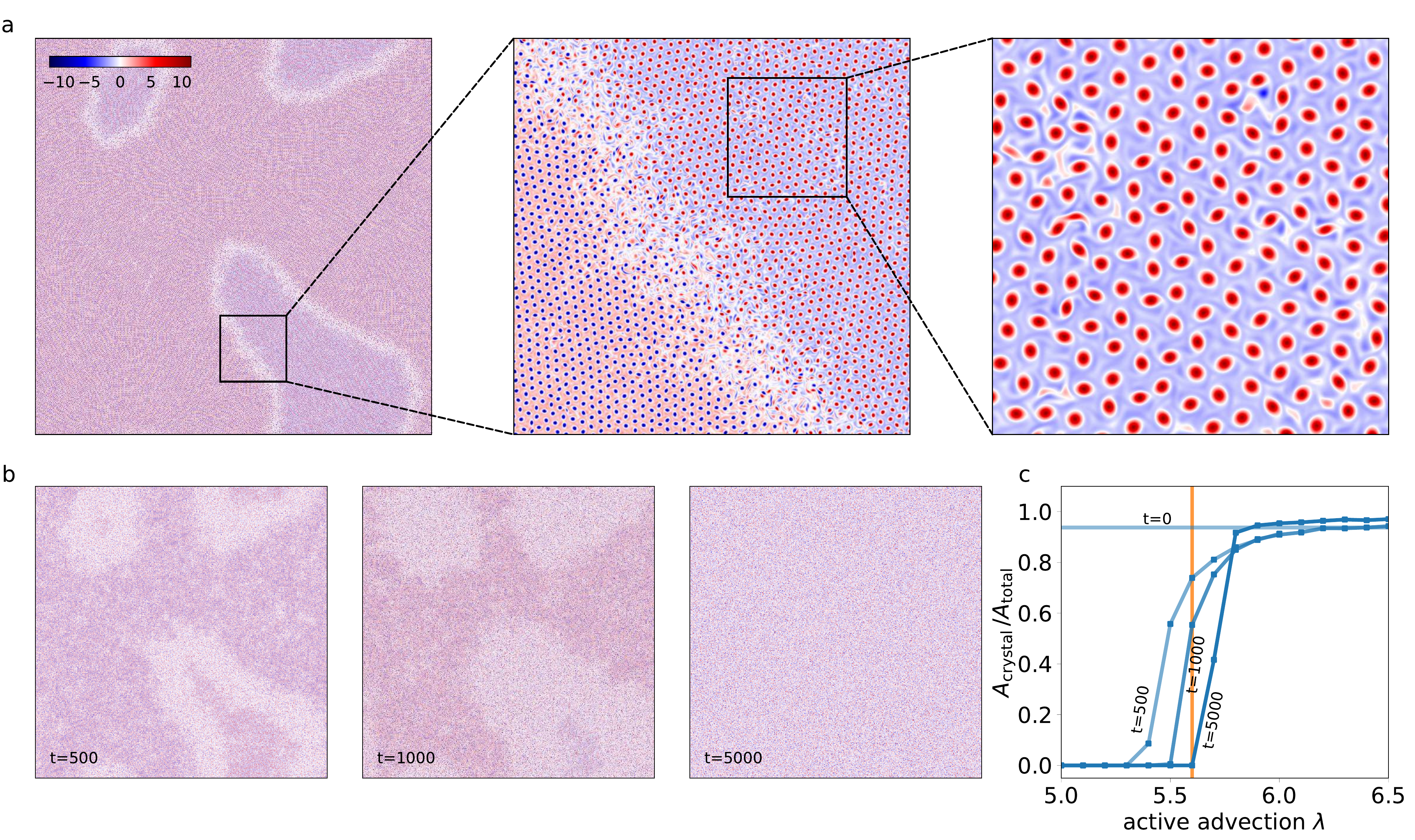}
\caption{
Active vortex crystals and their melting: (a)~Our active fluid simulation (solving \eqref{eq:equationofmotion} with $\lambda=7$, $\alpha=-0.8$ on a $1000\pi \times 1000\pi$ domain) exhibits AVC superstructures comprised of meta-stable opposite-polarity domains. The zoom-in shows that these domains are demarcated by an interfacial layer of turbulent active fluid. Panel (b) illustrates the melting of the AVC superstructures shown in panel (a) for $\lambda = 5.6$ after time $t=500$, $t=1000$, and $t=5000$. (c)~Area fraction of the crystal domains as a function of active advection for different times, starting from the superstructure~a at time $t=0$. Note that the transient width of the domain boundaries in the superstructures is controlled by the strength of the active advection. The orange line corresponds to the snapshots shown in panel (b).}
\label{fig:experiments}
\end{figure*}

In thermal equilibrium, 2D crystals can exhibit long-range orientational order while positional order is suppressed at finite temperatures~\cite{mermin1966absence,hohenberg1967existence,mermin1968crystalline,halperin2019jstat}. Over the last decades, several competing theories on the type and nature of solid-liquid transitions in 2D equilibrium systems have been proposed~\cite{strandburg1988two}.  Approaches such as density-functional~\cite{ramakrishnan1979first} and grain-boundary theories~\cite{chui1983grain} suggest possible mechanisms for first-order transitions. Seminal work by Kosterlitz, Thouless, Halperin, Nelson, and Young (KTHNY)~\cite{halperin1978theory} predicted a two-step continuous melting transition which proceeds through an intermediate hexatic phase characterized by quasi-long-range orientational order and short-ranged positional order. Such a hexatic phase has been observed in experiments on colloidal systems~\cite{zahn2000dynamic} and superconducting lattices~\cite{guillamon2009direct} as well as in numerical simulations of repulsive disks~\cite{kapfer2015two}.
\par
The complex melting dynamics of 2D equilibrium crystals raises fundamental questions about liquid-solid transitions in far-from-equilibrium systems~\cite{goldman2003lattice,boyer2009two,perlekar2010turbulence,paliwal2020role}. In particular, it remains an open question to which extent concepts from equilibrium melting processes translate to non-equilibrium melting. Thus far, experimental and numerical studies of particulate active matter draw a complex picture. Whereas Monte Carlo simulations for active particles with inverse-power-law repulsion~\cite{klamser2018thermodynamic}  reported an intermediate hexatic phase consistent with KTHNY scaling prediction, agent-based simulations~\cite{weber2014defect} and active Brownian particle simulations~\cite{paliwal2020role} suggest that active crystal structures can melt into a hexatic state without the KTHNY-typical unbinding of topological defect pairs.
\par
Active fluids are another important class of non-equilibrium systems, which show intriguing transitions from active turbulence~\cite{wensink2012meso,dunkel2013minimal,doostmohammadi2017onset,james2018vortex,james2018turbulence} to highly ordered vortex arrays. Vortex arrays have been observed in dense suspensions of swimming sperm cells~\cite{riedel2005self} or microtubule \cite{sumino2012large}, and have been have been predicted to form spontaneously by a wide range of generic active fluid models~\cite{james2018turbulence,dunkel2013minimal,doostmohammadi2017onset,shendruk2017dancing,2017SlomkaDunkel_PRF,Doostmohammadi:2016aa,2016OzHeDu_EPJE,nagai2015collective,grossmann2014vortex,dunkel2013minimal}. However, so far it has not been established whether such active systems exhibit true crystalline order, and if so, how active vortex crystals (AVCs) melt.

The systematic study of AVC formation and melting has remained an unsolved challenge owing to the prohibitively large system sizes and simulation times required. Overcoming previous limitations through direct active fluid simulations at unprecedented scale (Fig.~\ref{fig:experiments}a), we report here a detailed computational investigation of AVC emergence and melting in an experimentally validated generalized Toner-Tu model~\cite{wensink2012meso,bratanov2015new}. By evaluating the corresponding order parameters, we establish conclusively that the active vortex arrays indeed show true crystalline order. Our analysis furthermore predicts a rich spectrum of transition phenomena in AVCs, including hexatic phases,  hysteretic liquid-solid coexistence and metastable superstructures exhibiting spontaneously broken chiral symmetry. We also find that the emergence of AVCs shows intriguing transient features as a result of the self-organization of AVCs through a turbulent transient, followed by the slow coarsening dynamics of large crystal domains of opposite polarity.

\section*{Results}

\begin{figure*}
\centering
\includegraphics[width=0.75\textwidth]{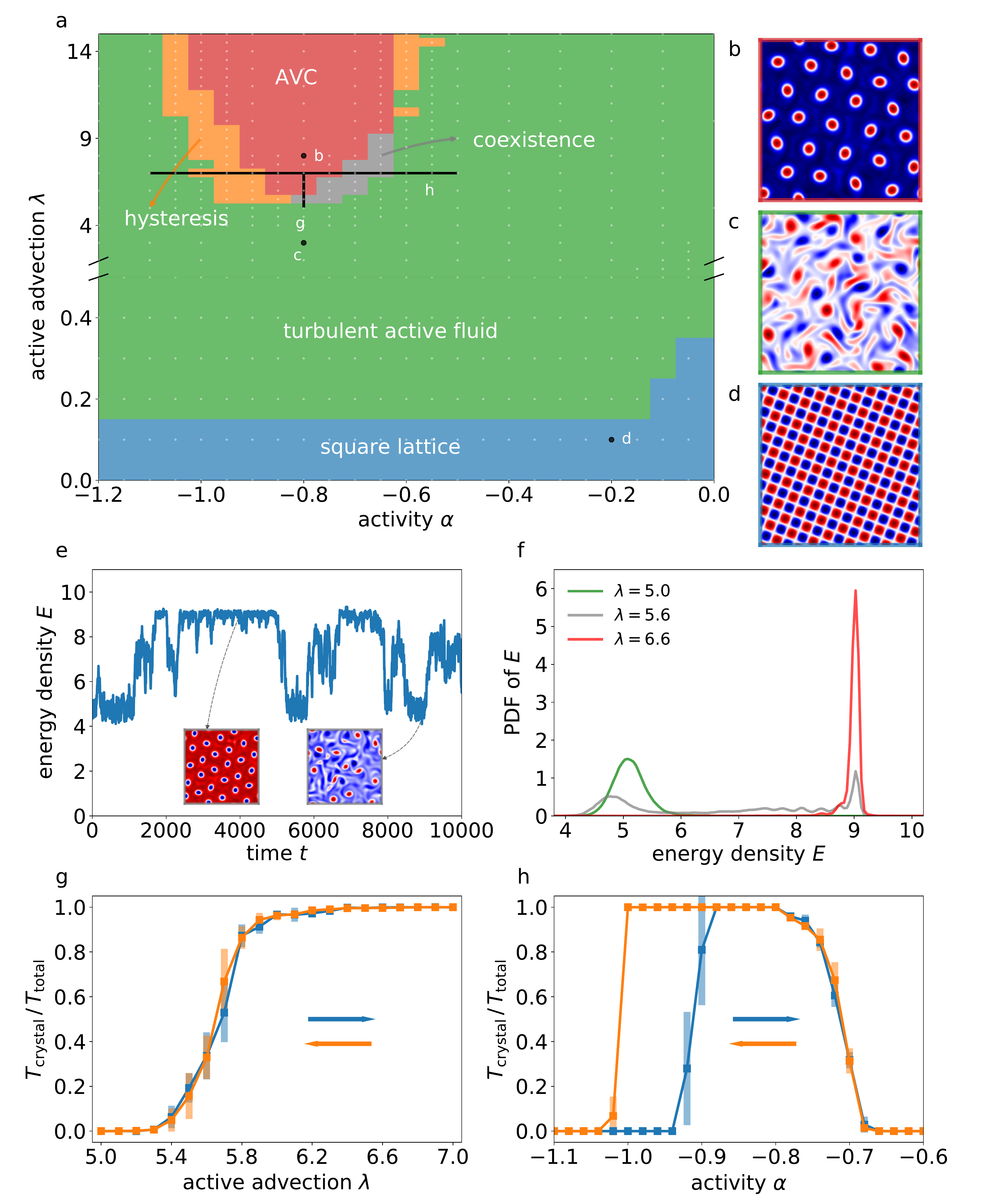}
\caption{ 
Phase diagram and melting transition ($L=20\pi$): (a) Different phases of the active matter system as a function of activity and active advection, obtained from close to 1000 simulations (Methods). Red, green, and blue regions correspond to (b) vortex crystal, (c) active fluid and  (d) square lattice, respectively. The gray and orange regions are the marginal stability regions between the active turbulence phase and the vortex crystal. The white dots show the parameter configuration used to obtain the phase diagram (see also Fig.~\ref{fig:meth_phase_diagram}). (e) A typical energy density time series for a simulation in the marginal stability region illustrates the intermittent melting and crystallization of the AVC. The insets show representative snapshots of the vorticity field. (f) Probability density functions of the energy density for values of $\lambda =$ 5 (green), 5.6 (gray) and 6.6 (red). (g) Melting transition of the AVC as a function of active advection ($\alpha=-0.8$) and (h) transition along the activity ($\lambda$=7) axis. The blue and orange curves correspond to increasing and decreasing values, respectively, of $\alpha$ and $\lambda$.}
\label{fig:phase_diagram}
\end{figure*}

 Our starting point is a generalization of the incompressible Toner-Tu equations~\cite{toner1998flocks,toner2005hydrodynamics,marchetti2013hydrodynamics} for the active fluid velocity field $\bs u$  and pressure field $p$~\cite{wensink2012meso,dunkel2013minimal,bratanov2015new}:
\begin{align}
  \partial_t \bs u + \lambda \bs u \cdot \nabla \bs u &= -\nabla p - (1+\Delta)^2 \bs u - (\alpha + \beta |{\bs u}|^2)\bs u,\nonumber\\
  \nabla\cdot {\bs u} &= 0,
  \label{eq:equationofmotion}
\end{align}
where $\lambda$ is an active advection parameter which incorporates the effects of active nematic stresses~\cite{dunkel2013minimal,2016HeEtAl_PRE,PhysRevE.97.022613}. The activity parameter $\alpha<0$  is proportional to the square of self-propulsion velocity, and $\beta$ sets the velocity relaxation time scale (which  can be scaled out). Active flows described by \eqref{eq:equationofmotion} are self-driven through a linear instability induced by the Swift-Hohenberg operator $(1+\Delta)^2$, which favors periodic flow patterns of wave length $2\pi$~\cite{simha2002hydrodynamic,saintillan2008instabilities}. \eqref{eq:equationofmotion} can be derived from a generic agent-based model~\cite{2016HeEtAl_PRE,PhysRevE.97.022613} that accounts for the particles' self-propulsion, their hydrodynamic interactions near a surface, and their steric interactions. It has been shown~\cite{wensink2012meso} that this minimal continuum theory quantitatively captures essential statistical properties of dense bacterial suspensions in quasi-2D microfluidic channels. Here, we apply \eqref{eq:equationofmotion} to study AVC dynamics by performing large-scale numerical simulation on a doubly periodic domain of size $L\times L$ with $L$ ranging up to $1000\pi$  (Fig.~\ref{fig:experiments}a), using a pseudo-spectral method for the spatial discretization and a fourth-order Runge-Kutta scheme for time stepping (Methods).

\subsection*{Large-scale active vortex crystals.}

As an illustration of the AVC emerging in our system, Fig.~\ref{fig:experiments}a shows the vorticity field for a simulation on a $1000\pi \times 1000\pi$ domain resolved with $8192^2$ grid points [\href{https://youtu.be/p_br0pkBEVs}{Supplementary Video~\ref{vid:superstructureevolution}}]. At this very large system size, AVC superstructures, i.e.~crystal domains of opposite polarity with domain boundaries comprised of active turbulence regions, crystallize from an initially turbulent active fluid. The highly dynamic domain boundaries play a prominent role in the melting of AVC superstructures, which can be induced, for example, by decreasing the active advection parameter. As active advection is decreased, the superstructures melt, and the turbulent boundary layers spread in area, destroying the crystal structure (Fig.~\ref{fig:experiments}b). Since the AVC superstructures are formed by crystal domains demarcated by an active fluid interfacial layer, a natural order parameter for our analysis is the fraction of area covered by the crystal domain $A_\mathrm{crystal}/A_\mathrm{total}$ (Methods). 
\par
To illustrate this transition, we evaluate the crystal area fraction as a function of active advection (at a fixed activity $\alpha=-0.8$) for different times, which is shown in Fig.~\ref{fig:experiments}c. Below $\lambda\approx 5.6$, the crystal domains melt completely into a statistically isotropic active fluid. Above a critical value of $\lambda=6.0$, almost the entire domain is covered by vortex crystals, with the area between the crystals of different polarity occupied by a layer of active fluid. There is a consistent, but slow decrease in the area of this boundary layer as advection is increased. In the following sections, we present a detailed characterization of the crystalline order and its melting.

\subsection*{Non-equilibrium phase diagram.}

We performed $\sim 1000$ simulations on smaller domains of size $L=20\pi$ to map out the $(\alpha,\lambda)$ phase diagram shown in Fig.~\ref{fig:phase_diagram}a.  
These parameter scans revealed three distinct phases:  The AVC phase Fig.~\ref{fig:phase_diagram}b forms  for $\alpha\approx -0.8$ and sufficiently large values of the active advection parameter $\lambda$, corresponding to strong extensile stresses (red-colored domain in Fig.~\ref{fig:phase_diagram}a). In this phase, vortices of same spontaneously chosen handedness self-organize in a triangular lattice, phenomenologically similar to those observed in dense sperm suspensions~\cite{riedel2005self}. Strikingly, this spontaneous symmetry breaking occurs after an initial turbulent transient [\href{https://youtu.be/n-2Yue-WFWc}{Supplementary Video \ref{vid:vortexarrayemergence}}]. The AVC phase is surrounded by an extended active turbulence (AT) phase (green domain in Fig.~\ref{fig:phase_diagram}a), in which transient vortices of either handedness coexist in the fluid (Fig.~\ref{fig:phase_diagram}c). Finally, for low active advection $\lambda\ll 1$, corresponding to contractile stresses~\cite{dunkel2013minimal}, the system settles into a stationary square flow-lattice state (Fig.~\ref{fig:phase_diagram}d), which can be explained with classical pattern formation theory~\cite{james2018turbulence}. We focus in the remainder on the \lq solid-liquid\rq{}  transitions between the AVC and AT phases, corresponding to the regions separating the red and green domains in Fig.~\ref{fig:phase_diagram}a. 
\par
Unexpectedly, our simulations reveal two distinct AVC-AT transition scenarios, characterized by  phase coexistence  and hysteresis, respectively. To demonstrate the characteristics of phase coexistence (gray in Fig.~\ref{fig:phase_diagram}a), we keep the activity parameter $\alpha=-0.8$ fixed and decrease the active advection parameter $\lambda$ (vertical scan). This transition is characterized by an intermittent switching between active turbulence and AVCs [\href{https://youtu.be/jQZlmNoSzFs}{Supplementary Video~\ref{vid:marginalstability}}]. The energy time series of a corresponding simulation is shown in Fig.~\ref{fig:phase_diagram}e. In the AVC state, the energy density is high due to the close packing of vortices, and the fluctuations are low. In the active fluid phase, the energy is lower and fluctuations are larger.  The energy-density probability density functions (PDFs) for three representative intermediate values of $\lambda$ along the vertical scan quantifies the relative abundance of each phase (Fig.~\ref{fig:phase_diagram}f). This temporal, highly dynamic phase coexistence is confirmed by measuring the fraction of time in the crystal phase, which is shown for the vertical scan in Fig.~\ref{fig:phase_diagram}g.
\par
To illustrate the hysteretic transition  (orange in Fig.~\ref{fig:phase_diagram}a), we keep the active advection fixed at value $\lambda=7$ and change the activity parameter $\alpha$ (horizontal scan). In the transition region for low activities, a vortex crystal will not emerge from random initial conditions for these parameters, but the crystal itself is a stable solution. This is illustrated by the AVC time fraction for the horizontal scan shown in Fig.~\ref{fig:phase_diagram}h, which clearly exhibits a hysteresis loop. As the activity parameter is further increased, a second transition without hysteresis is observed. Closer to the boundary of the active turbulence region, the vortex arrays start showing a fluid-like arrangement of vortices rather than crystal-like or hexatic. For a systematic characterization, we present results from significantly larger domains in the following sections.

\subsection*{Hexatic phase.}

\begin{figure}
\centering
\includegraphics[width=\columnwidth]{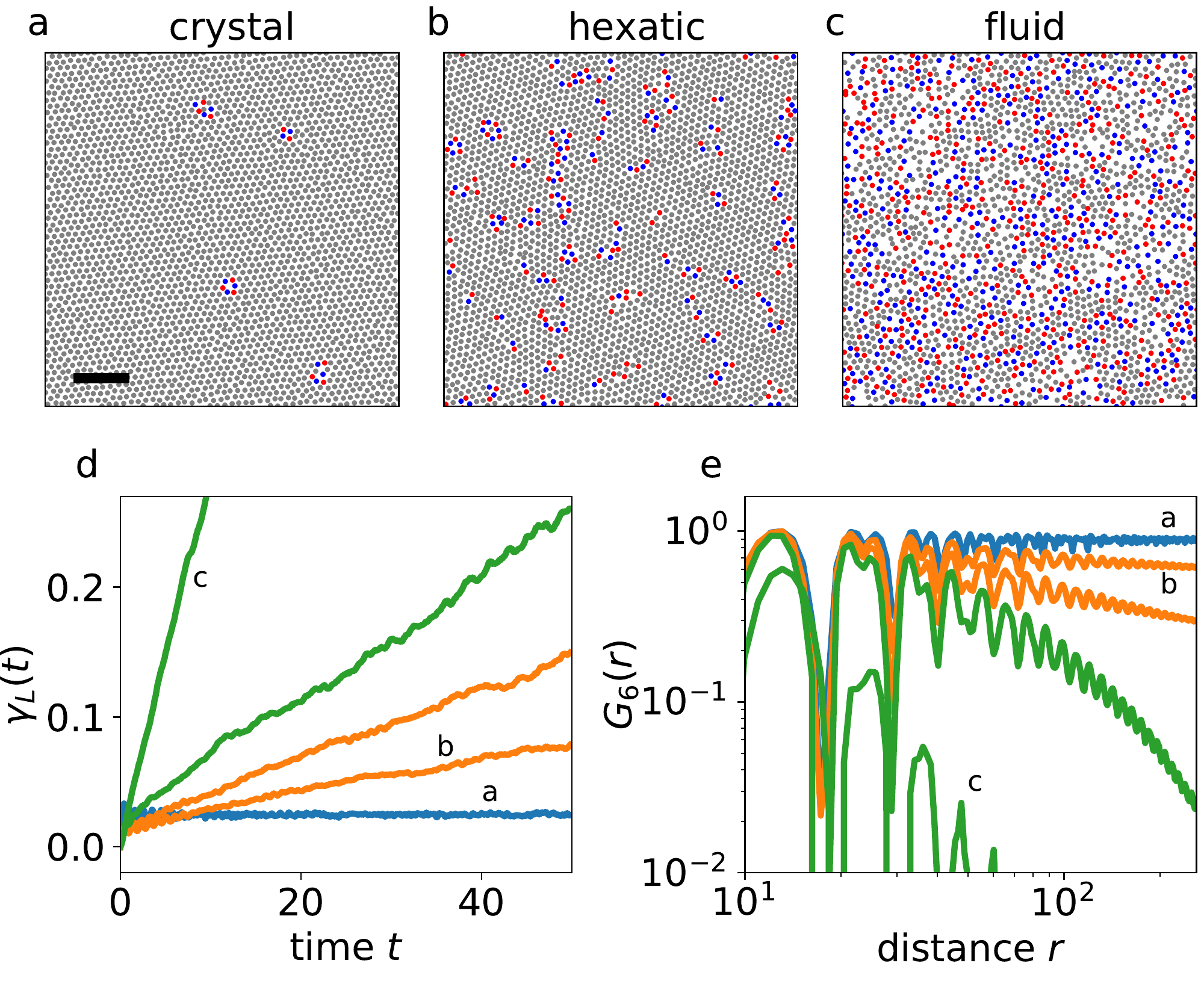}
\caption{Characterization of the melting transition: Vortex cores (gray) and the locations of the 5-fold (red) and 7-fold (blue) defects in the (a) crystal ($\alpha=-0.8,\ \lambda=15$), (b) hexatic ($\alpha=-0.75,\ \lambda=7$) and (c) fluid ($\alpha=-0.7,\ \lambda=7$) phases. The scale bar denotes $L=100$. (d) Dynamic Lindemann parameter $\gamma_L(t)$ as a function of time and (e) orientational correlation function $G_6(r)$ for different parameter choices (Methods) in the crystal (blue), hexatic (orange) and fluid (green) phases. $\gamma_L(t)$ remains bounded for the crystal phase, whereas it diverges for both the hexatic and fluid phases. $G_6(r)$, while remaining constant for the crystal, decay algebraically as it melts, demonstrating the existence of an intermediate hexatic phase. In the fluid phase, $G_6(r)$ decays faster than algebraic.}
\label{fig:phase_transition}
\end{figure}

To characterize the crystalline order in this system, we perform simulations that are two orders of magnitude larger than for the phase diagram ($L=200\pi$), and construct a Voronoi partition to identify 5-fold and 7-fold defects (Methods). Deep inside the crystal regime, we observe a well-ordered structure with bound pairs of dislocations (Fig.~\ref{fig:phase_transition}a: $\lambda=15, \alpha=-0.8$). Closer to the active turbulence region, the order starts unravelling (Fig.~\ref{fig:phase_transition}b: $\lambda=7, \alpha=-0.75$, and c: $\lambda=7, \alpha=-0.7$). In Fig.~\ref{fig:phase_transition}b, we observe the unbinding of dislocation pairs, as well as a few free disclinations. As we move closer to the active turbulence region, Fig.~\ref{fig:phase_transition}c, clusters of defects emerge.
As we demonstrate below, Fig.~\ref{fig:phase_transition}b corresponds to a hexatic phase, which points at a connection to equilibrium melting transitions following the KTHNY scenario.
\par
As a quantitative characterization of the crystalline phase we use the dynamic Lindemann parameter, which is defined as the relative displacement of neighboring vortex cores~\cite{bedanov1985modified,zahn1999two,zahn2000dynamic}:
\begin{equation} \label{eq:lindemann}
\gamma_L(t)=\left\langle(\Delta{\boldsymbol{x}}_i(t)-\Delta{\boldsymbol{x}}_{i+1}(t))^2\right\rangle/2a^2.
\end{equation}
Here, $\Delta \boldsymbol{x}_i(t)=\boldsymbol{x}_i(t)-\boldsymbol{x}_i(0)$ is the temporal displacement of a vortex core position $\boldsymbol{x}_i(t)$ from its initial position $\boldsymbol{x}_i(0)$, $i$ and $i+1$ denote neighbors, and $a$ is the lattice spacing. For a crystal, $\gamma_L(t)$ remains bounded whereas for both hexatic and fluid phases, it diverges with time. Fig.~\ref{fig:phase_transition}d shows that the dynamic Lindemann parameter for our system remains bounded well inside the AVC regime. Close to the transition region, $\gamma_L(t)$ diverges, indicating either a hexatic or a fluid phase.
\par
To clearly distinguish the crystalline, hexatic and fluid phase, respectively, we evaluate the orientational correlation function $G_6(r)$ ~\cite{strandburg1988two}. To this end, we calculate the orientational order $\psi_i=\sum_j \mathrm{exp}(6\mathrm{i}\theta_{ij})/N(i)$ for each lattice site $i$. Here, $\theta_{ij}$ is the angle between the line connecting the neighbors $i$ and $j$ and an arbitrary axis, and $N(i)$ is the number of neighbors. The orientational correlation is then defined as
\begin{equation}\label{eq:G6}
G_6(r)=\left\langle\psi_i^*\psi_j\delta(r-r_{ij})\right\rangle/\left\langle\delta(r-r_{ij})\right\rangle
\end{equation} where $r_{ij}$ is the distance between vortex cores $i$ and $j$, and the average is over all lattice sites $i$ and $j$. Fig.~\ref{fig:phase_transition}e shows the orientational correlation function for different regions in the phase diagram. For the crystal phase, as expected, we observe long-range orientational order. Closer to the transition region, $G_6(r)$ shows quasi-long-range order characterized by an algebraic decay. Finally, for the fluid phase, the orientational correlation function decays faster than algebraic. This indicates that this AVC-AT transition proceeds through an intermediate hexatic phase, which establishes a profound analogy to melting transitions in 2D equilibrium systems.

\subsection*{Emergence of AVCs.}

\begin{figure}
\centering
\includegraphics[width=0.9\columnwidth]{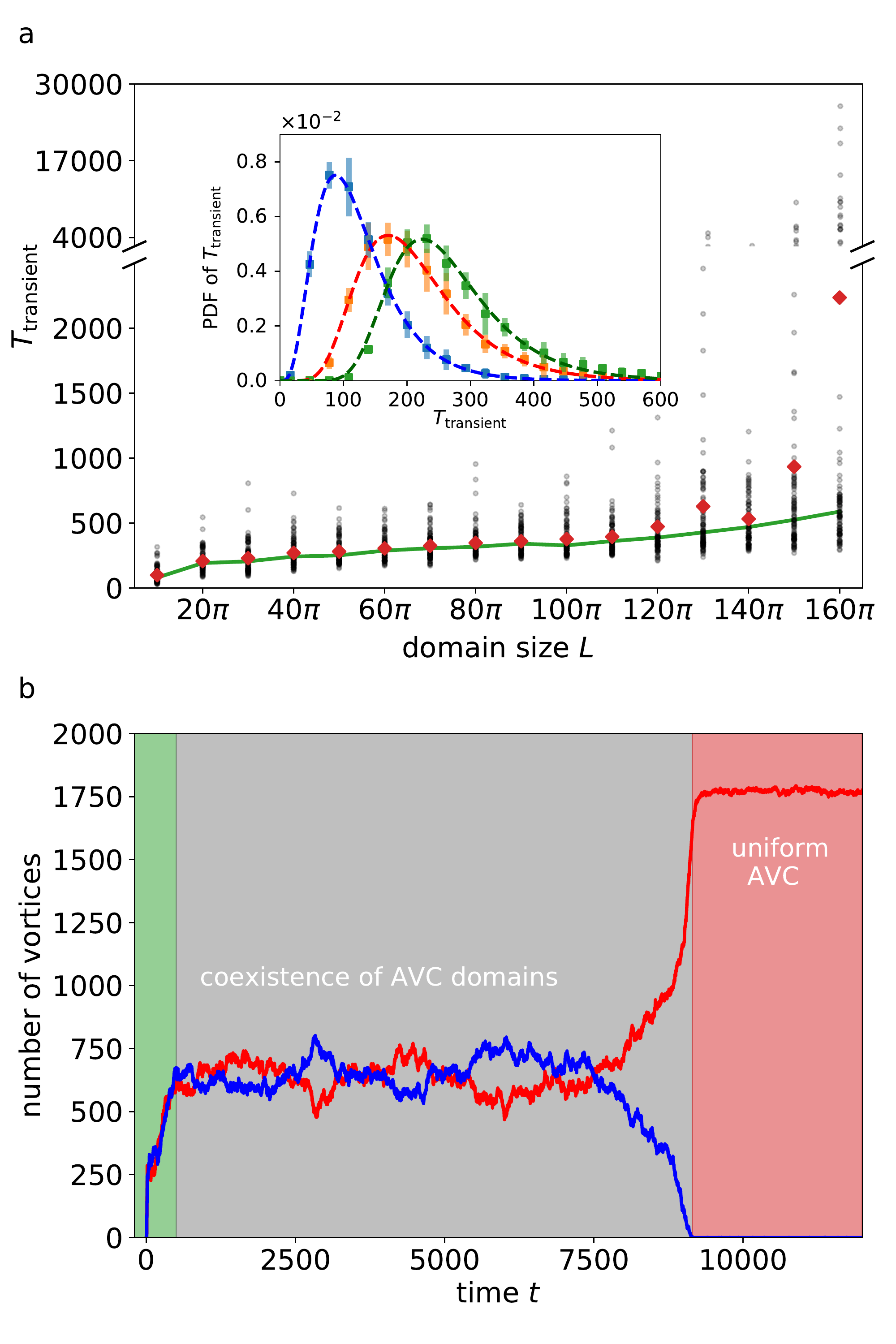}
\caption{Transient durations of AVC emergence: (a) Duration of the transients leading to a uniform AVC domain as a function of domain size $L$. The red dots and the green curve are the mean and median values, respectively. The change in slope at about $L=120\pi$ marks the domain size where AVC superstructures start to become stable. (inset) Probability density function of the transient duration for $L= 10\pi$ (blue), $20\pi$ (orange) and $40\pi$ (green), obtained from $10^4$ simulations each, and the corresponding fit with the theoretically proposed PDF \eqref{eq:transient_pdf} (dashed curves). (b) The time series of the number of positive (red) and negative vortices (blue) for a simulation with large domain size ($L=160\pi$). The green, gray, and red regions denote the initial transient, the coexistence of AVC domains of opposite polarity, and the final uniform AVC respectively.}
\label{fig:transients}
\end{figure}

Next, we characterize the emergence of AVCs as a function of system size.
To this end, we have determined the transient time until a uniform AVC is formed for an ensemble of 100 simulations for each system size, covering domain sizes between $L=10\pi$ and $L=160\pi$. Fig.~\ref{fig:transients}a shows the resulting scatter plot, which demonstrates that the lifetime of the transient active fluid depends sensitively on the initial condition and increases considerably with domain size. 
\par
For small domains, this is mainly rooted in the fact that the emergence of a uniform AVC occurs after a spontaneous discrete symmetry breaking through a turbulent transient~\cite{james2018turbulence}, which renders the transient time a random variable. In fact, the PDF of transition times is well captured by
\begin{equation} \label{eq:transient_pdf}
  P(T) = \frac{\delta}{\tau}\left[ 1-\mathrm{e}^{-\frac{T}{\tau}} \right]^{\delta-1}\mathrm{e}^{-\frac{T}{\tau}},
\end{equation}
where $\tau$ and $\delta$ depend on the vortex lifetime and domain size, respectively. This expression can be rationalized from the observation that vortex lifetimes in active turbulence have an approximately exponential distribution~\cite{james2018vortex}. A good estimate for the transient time is the time after which the spontaneous symmetry breaking occurs. Its distribution can be obtained from the PDF of the time it takes for one polarity of vortices to decay, which amounts to computing the maximum survival time of a set of like-signed vortices. Assuming statistical independence of the individual decay processes yields the proposed PDF \eqref{eq:transient_pdf}. Figure~\ref{fig:transients} (inset) shows the corresponding fits for the PDFs of the transient durations obtained from $10^4$ simulations for each domain size, demonstrating an excellent agreement.
\par
For sufficiently large domain sizes, an additional effect comes into play: crystal domains with both polarity can coexist, leading to metastable AVC superstructures with very long lifetimes (see, e.g., Fig.~\ref{fig:experiments}a). The temporal evolution of AVC clusters is illustrated in Fig.~\ref{fig:transients}b, which shows the number of positive and negative vortices as a function of time ($L=160\pi$). In this example, two vortex clusters of approximately equal sizes but opposite polarity coexist for more than 8000 nondimensional time units, before a uniform AVC forms. These metastable AVC superstructures explain the extreme outliers in the transient duration which are the cause for the sharp increase of the mean transient time for system sizes beyond $L=120\pi$. This divergence of transient durations with domain size, a supertransient feature, is observed in a variety of dynamical systems such as coupled map lattices and reaction diffusion systems~\cite{kaneko1990supertransients,crutchfield1988attractors,strain1998size,lai2011transient}. Unlike supertransient chaos, our transient state is characterized initially by a turbulent regime and then a gradually evolving superstructure. An extreme example of such a superstructure is shown in Fig.~\ref{fig:experiments}a for a domain of size $L=1000\pi$. The slowly evolving crystal domains are separated by a highly dynamic interfacial area of active turbulence [\href{https://youtu.be/p_br0pkBEVs}{Supplementary Video~\ref{vid:superstructureevolution}}].

\section*{Discussion}

Using large-scale simulations at unprecedented scale, we have determined the comprehensive phase diagram of a generalized Toner-Tu model for active fluids. Our analysis establishes that the self-organized active vortex arrays show crystalline order and predicts a rich melting scenario from active vortex crystals to active turbulence. Depending on the path through the parameter space, we find two distinct transition scenarios. The first transition scenario resembles a KTHNY-like melting. When transitioning from the AVC phase to the AT phase, we first observe the unbinding of defect pairs, which results in the loss of long-range orientational order before melting into an active fluid. In this parameter range, small domains exhibit a highly dynamic melting and re-crystallization. On very large domains, broken-symmetry AVC domains of opposite polarity emerge, whose melting results from the spreading of the turbulent interfacial layers. Intriguingly, in contrast to the KTHNY-like melting, the second transition scenario shows hysteresis between the AVC and the AT state. Generally, our results reveal the roles of nonlinear advection and activity in the complex self-assembly and melting of AVCs, and highlight deep connections between phase transitions in active matter and their classical equilibrium counterparts.
\par
Experimental tests of our predictions could be an exciting direction for future work.
Dense suspension of spermatozoa, which show both, active turbulence~\cite{creppy2015turbulence} as well as self-organized regular vortex arrays~\cite{riedel2005self} are arguably the best candidate to test our prediction on AVC-AT transitions. Although previous experiments~\cite{riedel2005self} so far suggest a fluid-like arrangement of vortices, rather than crystal-like or hexatic, it may be possible to achieve long-range crystal order through a careful tuning of experimental conditions~\cite{fisher2014dynamics}. For instance, the type of sperm cells as well as the intracellular ionic concentrations could affect the nature of sperm motility~\cite{alvarez2012rate}. Furthermore, the analysis of crystalline order in such systems would also require conducting experiments on large domains. If a vortex crystal phase is achieved, the activity can be tuned, for instance, by changing the motility through the ambient temperature~\cite{alavi2005sperm} to induce a potential melting transition. The preferred handedness of sperm cells on planar surfaces~\cite{woolley2003motility} precludes the observation of a spontaneously broken chiral symmetry. This could be alleviated by confinement between two walls. Such experiments will significantly enhance our knowledge of crystalline order, not just in active matter, but in out-of-equilibrium systems in general.

\section*{Methods}

\subsection*{Simulation details.}

We perform direct numerical simulations of the vorticity field $\omega=\nabla\times\bs u$ in a periodic domain by using a fully dealiased pseudo-spectral algorithm. The mean velocity $\langle\bs u\rangle$ is integrated separately. The corresponding evolution equations follow from \eqref{eq:equationofmotion} and take the form:
\begin{align}
\partial_t\omega + \lambda \bs u\cdot\nabla \omega &= -\big(1+\Delta\big)^2\omega -\alpha\omega -\beta\nabla\times\left(\left|\bs u\right|^2\,\bs u\right) 
\label{eq:vorticity_eom}\\
\partial_t\langle\bs u\rangle &= -\big(1+\alpha\big)\langle\bs u\rangle -\beta\left\langle\left|\bs u\right|^2\,\bs u\right\rangle. 
\label{eq:mean_u}
\end{align}
We solve \eqref{eq:vorticity_eom} and \eqref{eq:mean_u} with a fourth-order Runge-Kutta method for time stepping combined with an integrating factor for the linear terms. Our code is parallelized using GPUs (graphics processing units) in order to accelerate the computations. For the results discussed in the main text, the parameter values are listed in Table~\ref{tab:methods}.

\begin{table*}
\centering
\def\arraystretch{1.4}
\setlength{\tabcolsep}{1em}
\begin{tabular}{lccccc}
Figure	& $L$ & $\lambda$	&  $\alpha$ &$N$	&$\Delta t$\\
\hline
\ref{fig:experiments}a &$1000\pi$ &$7.0$ &$-0.800$ &$8192^2$ &$0.005$\\
\ref{fig:experiments}b &$1000\pi$ &$5.7$ &$-0.800$ &$8192^2$ &$0.005$\\
\ref{fig:experiments}a &$1000\pi$ &$[5,6.5]$ &$-0.800$ &$8192^2$ &$0.005$\\
\ref{fig:phase_diagram}a &$20\pi$ &$[0.0,\,15.0]$ &$[-1.200,\,0.000)$ &$256^2$ &$0.005$\\
\ref{fig:phase_diagram}b &$20\pi$ &$8.0$ &$-0.800$ &$1024^2$ &$0.001$\\
\ref{fig:phase_diagram}c &$20\pi$ &$3.0$ &$-0.800$ &$1024^2$ &$0.001$\\
\ref{fig:phase_diagram}d &$20\pi$ &$0.1$ &$-0.200$ &$1024^2$ &$0.001$\\
\ref{fig:phase_diagram}e &$20\pi$ &$5.6$ &$-0.800$ &$256^2$ &$0.005$\\
\ref{fig:phase_diagram}f &$20\pi$ &$5.0,\,5.6,\,6.6$ &$-0.800$ &$256^2$ &$0.005$\\
\ref{fig:phase_diagram}g &$20\pi$ &$[5.0,\,7.0]$ &$-0.800$ &$256^2$ &$0.005$\\
\ref{fig:phase_diagram}h &$20\pi$ &$7.0$ &$[-1.100,\,-0.600]$ &$256^2$ &$0.005$\\
\ref{fig:phase_transition}a & $200\pi$ & $15.0$ & $-0.800$ & $1024^2$ & $0.005$\\
\ref{fig:phase_transition}b & $200\pi$ & 7.0 & -0.750 & $1024^2$ & $0.005$\\
\ref{fig:phase_transition}c & $200\pi$ & 7.0 & -0.700 & $1024^2$ & $0.005$\\
\ref{fig:phase_transition}d & $200\pi$ & 7.0 & -0.750,\,-0.725,\,-0.715,\,-0.700 &$1024^2$ & $0.005$\\
\ref{fig:phase_transition}d & $200\pi$ & 15.0 & -0.800 &$1024^2$ & $0.005$\\
\ref{fig:phase_transition}e & $200\pi$ & 7.0 & -0.750,\,-0.725,\,-0.715,\,-0.700 &$1024^2$ & $0.005$\\
\ref{fig:phase_transition}e & $200\pi$ & 15.0 & -0.800 &$1024^2$ & $0.005$\\
\ref{fig:transients}a &$[10\pi,\,160\pi]$ &$7.0$ &$-0.800$ &$256^2$, $512^2$, $1024^2$ &$0.005$\\
\ref{fig:transients}a inset &$10\pi$, $20\pi$, $40\pi$ &$7.0$ &$-0.800$ &$256^2$ &$0.005$\\
\ref{fig:transients}b &$160\pi$ &$7.0$ &$-0.800$ &$1024^2$ &$0.005$\\
\ref{fig:boundary_layer_methods} &$1000\pi$ &$5.6$ &$-0.800$&$8192^2$ &$0.005$\\
\end{tabular}
\caption{Simulation parameters: Domain size $L$, active advection parameter $\lambda$, activity parameter $\alpha$, number of grid points $N$, time step $\Delta t$. The parameter $\beta$ is set to $0.01$ in all simulations.}
\label{tab:methods}
\end{table*}

\subsection*{Phase diagram.}

\begin{figure}
	\centering
	\includegraphics[width=1.0\columnwidth]{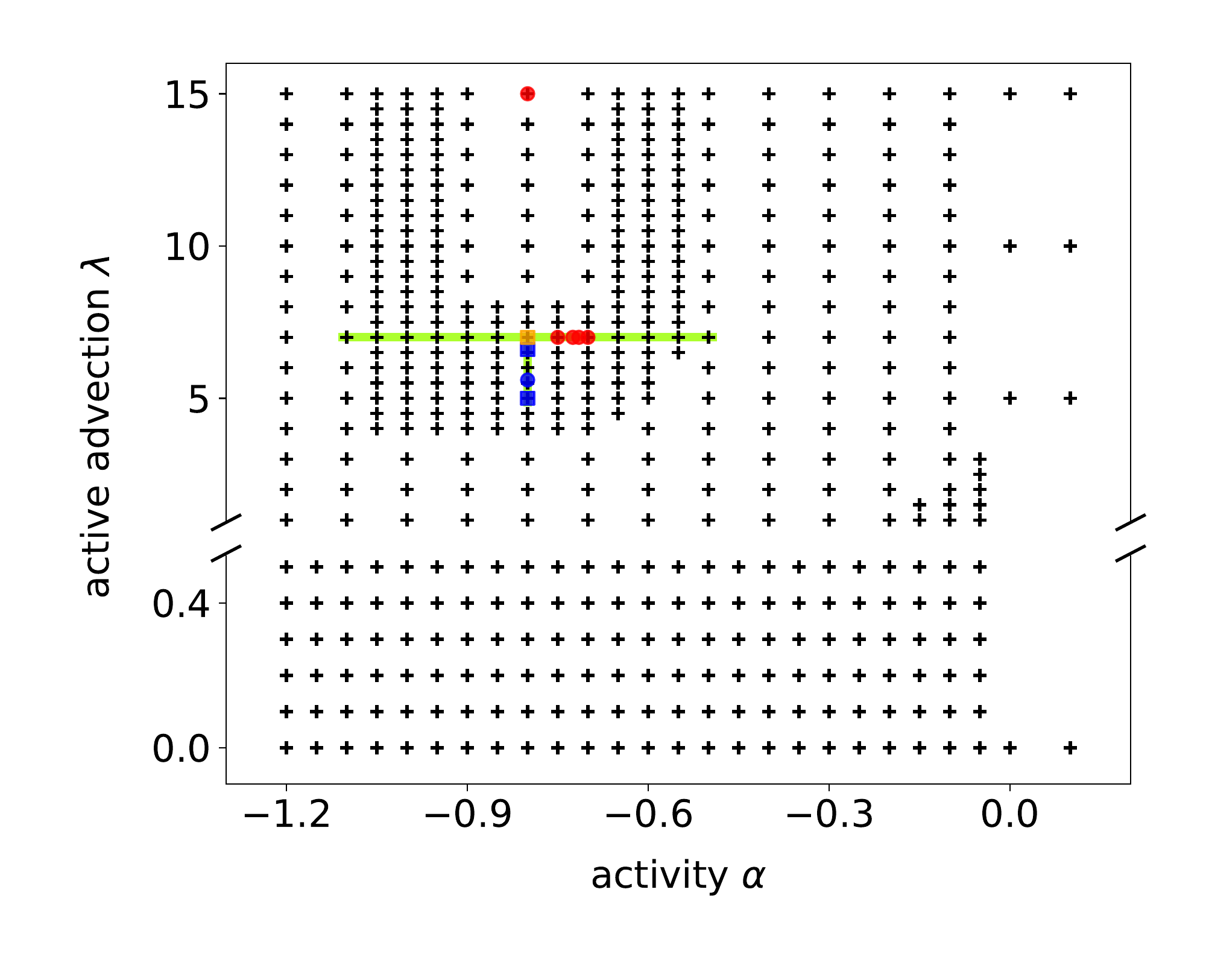}
	\caption{Overview of simulation parameters: The crosses show the parameter configuration used to obtain the phase diagram (Fig.~\ref{fig:phase_diagram}, domain size $L=20\pi$). Each cross  represents simulations with two different initial conditions, as noted in the main text. The PDFs in Fig.~\ref{fig:phase_diagram}f are based on simulations with parameters marked in blue ($L=20\pi$). The phase transition curves Fig.~\ref{fig:phase_diagram}g and h are based on simulations with parameters marked in green ($L=20\pi$). The parameters used in Fig.~\ref{fig:phase_transition} are indicated in red ($L=200\pi$). Fig.~\ref{fig:transients} ($L=10\pi-160\pi$) and \ref{fig:experiments}a ($L=1000\pi$) are based on the parameter choice shown in yellow. Fig.~\ref{fig:experiments}b ($L=1000\pi$) and \ref{fig:phase_diagram}e ($L=20\pi$) are based on the parameter choice indicated by a blue circle. See also Table~\ref{tab:methods} for the parameter values.}
	\label{fig:meth_phase_diagram}
\end{figure}

The phase diagram (Fig.~\ref{fig:phase_diagram}a) is obtained from simulations of 477 different parameter configurations as shown in Fig.~\ref{fig:meth_phase_diagram}. For each configuration, we use two different initial conditions: a random initial condition and a vortex crystal.

The different phases shown in Fig.~\ref{fig:phase_diagram}a are defined as follows. The square lattice, active turbulence and vortex crystal phases show obvious qualitative differences as noted in the main text and are easily distinguished visually. The hysteresis phase in the marginal stability region is identified as such when the simulations are bistable; the simulations starting with random initial conditions result in an active turbulence phase whereas a vortex crystal initial condition remains stable. The simulations are checked for convergence until a total simulation time of $T=2000$ ($4\times10^5$ time steps). The coexistence region is defined by evaluating the PDF of the energy density. If the PDF has two peaks (see, e.g., Fig.~\ref{fig:phase_diagram}f), it is defined as a temporally intermittent pattern.

\subsection*{Phase transition.} 

The phase transition between active turbulence and vortex crystals in small domains (Fig.~\ref{fig:phase_diagram}g and h) is characterized as follows. For the transition curves in both increasing and decreasing directions of parameter values, we conduct simulations in the range $5.0\leq\lambda\leq 7.0$ and $-1.1\leq\alpha\leq -0.6$. For $\lambda=5$ and $\alpha=-1.1$, we start our simulation from random initial conditions. For the rest of the simulations, the final snapshot of the previous simulation is used as the initial condition. Once a statistically steady state is reached (after about $10^6$ time steps), we collect data for $10^6$ time steps and evaluate the PDF of the energy density. If the PDF has only one peak, the order parameter $T_{\mathrm{crystal}}/T_{\mathrm{total}}$ takes the value $0$ or $1$, depending on the phase. Otherwise, the energy density at the minimum between the two peaks of the PDF, $E_{\mathrm{min}}$ is evaluated. The order parameter then takes the value of the probability that the energy density $E>E_{\mathrm{min}}$. This process is repeated five times, and the mean and the standard deviations are used to construct the transition curves and estimate the uncertainties,  which are shown in Fig.~\ref{fig:phase_diagram}g and h.

\begin{figure}
\centering
\includegraphics[width=1.0\columnwidth]{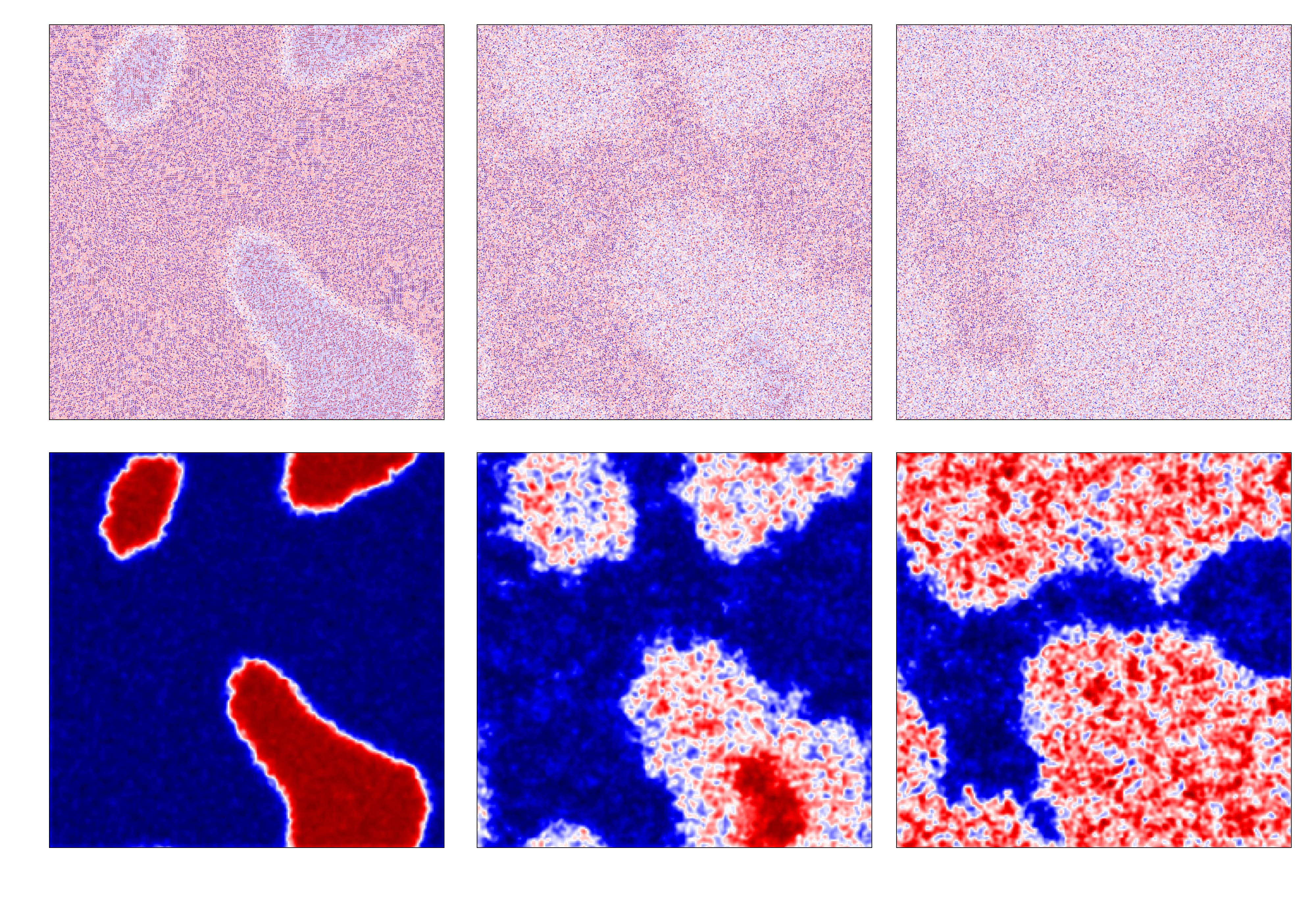}
\caption{Analysis of melting of AVC superstructures: Top row shows the vorticity field for $\lambda=5.6$ at different times (initial condition, at $t=1000$ and at $t=2000$). Bottom row shows the corresponding smoothed fields.}
\label{fig:boundary_layer_methods}
\end{figure}

To evaluate the phase transition curve in large domains (Fig.~\ref{fig:experiments}g), we first identify centers of the strong vortices~\cite{james2018vortex,giomi2015geometry}. Then an order parameter field is obtained by calculating, for each point $(x,y)$, the difference between the number of positive and negative vortices within a circle of radius $r$ centered at $(x,y)$ ($r$ is about half the mean distance between the nearest neighbors and next nearest neighbors). The resulting field is then smoothed using a Gaussian filter with standard deviation $\sigma=12$. The original vorticity field and the smoothed field are shown in Fig.~\ref{fig:boundary_layer_methods}. The turbulent region is then defined as the area where the absolute value of this smoothed field is less than half the maximum value of the field. Once this turbulent region is defined, the order parameter $A_{\mathrm{crystal}}/A_{\mathrm{total}}$ is calculated by evaluating the fraction of the total area covered by the vortex crystal.

\subsection*{Transient durations.} 

To evaluate transient durations (Fig.~\ref{fig:transients}a and b), we conduct simulations starting with random initial conditions until a converged vortex crystal state is reached for each domain size. The convergence is defined as follows. By employing a vortex identification algorithm~\cite{james2018vortex,giomi2015geometry}, we obtain a time series of the number of strong vortices of both polarity. A converged vortex crystal is obtained when the number of vortices of either sign reaches $93\%$ of the theoretical maximum number of vortices possible in the domain. To obtain the mean and median transient durations in Fig.~\ref{fig:transients}a, the simulations are repeated $100$ times for each domain size and the corresponding mean and median durations are calculated. The PDFs (inset Fig.~\ref{fig:transients}a) are obtained by evaluating the transient durations for three different domain sizes from $10^4$ simulations each, starting from random initial conditions. The error bars correspond to the difference between the maximum and the minimum from the 5 bootstrapped PDFs obtained from $2000$ simulations each. The theoretical curves are obtained by fitting \eqref{eq:transient_pdf} to the numerical data. The corresponding values of the free parameters $\delta$ and $\tau$ are, respectively, $4.97$ and $54.59$ for $L=10\pi$ (blue curve), $10.40$ and $72.78$ for $L=20\pi$ (orange curve) and $21.67$ and $72.81$ for $L=40\pi$ (green curve).

\subsection*{Defects, dynamic Lindemann parameter and orientational correlation.}

For each parameter choice in Fig.~\ref{fig:phase_transition}, simulations starting from random initial conditions are run for at least $1.2\times10^7$ time steps until a statistically steady state is reached. To identify the defects, centers of strong vortices are first evaluated~\cite{james2018vortex,giomi2015geometry}. By constructing a Voronoi partition (using the Python open source module scipy.spatial.Voronoi) over this vortex core configuration, 5-fold and 7-fold defects are determined~\cite{paliwal2020role}.

To evaluate the dynamic Lindemann parameter, we obtain $500$ snapshots separated by $\Delta t=0.1$, after the simulations have reached a converged state. For each snapshot, we identify the centers of strong vortices. The trajectory of each vortex core is then tracked. Only vortices which survive the entire duration of the simulation after reaching the statistically steady state are included in the analysis. The dynamic Lindemann parameter is then evaluated following \eqref{eq:lindemann} in the main text.
\par
The orientational correlation is evaluated according to \eqref{eq:G6} in the main text, after identifying the centers of the strong vortices in converged simulations. The results are averaged across $10$ snapshots (with $\Delta t=2000$) each from three simulations with different initial conditions.

\bibliography{main}

\begin{thebibliography}{10}
\expandafter\ifx\csname url\endcsname\relax
  \def\url#1{\texttt{#1}}\fi
\expandafter\ifx\csname urlprefix\endcsname\relax\def\urlprefix{URL }\fi
\providecommand{\bibinfo}[2]{#2}
\providecommand{\eprint}[2][]{\url{#2}}

\bibitem{strandburg1988two}
\bibinfo{author}{Strandburg, K.~J.}
\newblock \bibinfo{title}{Two-dimensional melting}.
\newblock \emph{\bibinfo{journal}{Rev. Mod. Phys.}}
  \textbf{\bibinfo{volume}{60}}, \bibinfo{pages}{161} (\bibinfo{year}{1988}).

\bibitem{palacci2013living}
\bibinfo{author}{Palacci, J.}, \bibinfo{author}{Sacanna, S.},
  \bibinfo{author}{Steinberg, A.~P.}, \bibinfo{author}{Pine, D.~J.} \&
  \bibinfo{author}{Chaikin, P.~M.}
\newblock \bibinfo{title}{Living crystals of light-activated colloidal
  surfers}.
\newblock \emph{\bibinfo{journal}{Science}} \textbf{\bibinfo{volume}{339}},
  \bibinfo{pages}{936--940} (\bibinfo{year}{2013}).

\bibitem{mognetti2013living}
\bibinfo{author}{Mognetti, B.~M.} \emph{et~al.}
\newblock \bibinfo{title}{Living clusters and crystals from low-density
  suspensions of active colloids}.
\newblock \emph{\bibinfo{journal}{Phys. Rev. Lett.}}
  \textbf{\bibinfo{volume}{111}}, \bibinfo{pages}{245702}
  (\bibinfo{year}{2013}).

\bibitem{klamser2018thermodynamic}
\bibinfo{author}{Klamser, J.~U.}, \bibinfo{author}{Kapfer, S.~C.} \&
  \bibinfo{author}{Krauth, W.}
\newblock \bibinfo{title}{Thermodynamic phases in two-dimensional active
  matter}.
\newblock \emph{\bibinfo{journal}{Nat. Commun.}} \textbf{\bibinfo{volume}{9}},
  \bibinfo{pages}{5045} (\bibinfo{year}{2018}).

\bibitem{weber2014defect}
\bibinfo{author}{Weber, C.~A.}, \bibinfo{author}{Bock, C.} \&
  \bibinfo{author}{Frey, E.}
\newblock \bibinfo{title}{Defect-mediated phase transitions in active soft
  matter}.
\newblock \emph{\bibinfo{journal}{Phys. Rev. Lett.}}
  \textbf{\bibinfo{volume}{112}}, \bibinfo{pages}{168301}
  (\bibinfo{year}{2014}).

\bibitem{paliwal2020role}
\bibinfo{author}{Paliwal, S.} \& \bibinfo{author}{Dijkstra, M.}
\newblock \bibinfo{title}{Role of topological defects in the two-stage melting
  and elastic behavior of active {B}rownian particles}.
\newblock \emph{\bibinfo{journal}{Phys. Rev. Research}}
  \textbf{\bibinfo{volume}{2}}, \bibinfo{pages}{012013} (\bibinfo{year}{2020}).

\bibitem{petroff2015fast}
\bibinfo{author}{Petroff, A.~P.}, \bibinfo{author}{Wu, X.-L.} \&
  \bibinfo{author}{Libchaber, A.}
\newblock \bibinfo{title}{Fast-moving bacteria self-organize into active
  two-dimensional crystals of rotating cells}.
\newblock \emph{\bibinfo{journal}{Phys. Rev. Lett.}}
  \textbf{\bibinfo{volume}{114}}, \bibinfo{pages}{158102}
  (\bibinfo{year}{2015}).

\bibitem{DUTTA2019743}
\bibinfo{author}{Dutta, S.}, \bibinfo{author}{Djabrayan, N. J.-V.},
  \bibinfo{author}{T., S.}, \bibinfo{author}{Shvartsman, S.~Y.} \&
  \bibinfo{author}{Krajnc, M.}
\newblock \bibinfo{title}{Self-similar dynamics of nuclear packing in the early
  drosophila embryo}.
\newblock \emph{\bibinfo{journal}{Biophys. J.}} \textbf{\bibinfo{volume}{117}},
  \bibinfo{pages}{743--750} (\bibinfo{year}{2019}).

\bibitem{riedel2005self}
\bibinfo{author}{Riedel, I.~H.}, \bibinfo{author}{Kruse, K.} \&
  \bibinfo{author}{Howard, J.}
\newblock \bibinfo{title}{A self-organized vortex array of hydrodynamically
  entrained sperm cells}.
\newblock \emph{\bibinfo{journal}{Science}} \textbf{\bibinfo{volume}{309}},
  \bibinfo{pages}{300--303} (\bibinfo{year}{2005}).

\bibitem{theurkauff2012dynamic}
\bibinfo{author}{Theurkauff, I.}, \bibinfo{author}{Cottin-Bizonne, C.},
  \bibinfo{author}{Palacci, J.}, \bibinfo{author}{Ybert, C.} \&
  \bibinfo{author}{Bocquet, L.}
\newblock \bibinfo{title}{Dynamic clustering in active colloidal suspensions
  with chemical signaling}.
\newblock \emph{\bibinfo{journal}{Phys. Rev. Lett.}}
  \textbf{\bibinfo{volume}{108}}, \bibinfo{pages}{268303}
  (\bibinfo{year}{2012}).

\bibitem{takatori2016acoustic}
\bibinfo{author}{Takatori, S.~C.}, \bibinfo{author}{De~Dier, R.},
  \bibinfo{author}{Vermant, J.} \& \bibinfo{author}{Brady, J.~F.}
\newblock \bibinfo{title}{Acoustic trapping of active matter}.
\newblock \emph{\bibinfo{journal}{Nat. Commun.}} \textbf{\bibinfo{volume}{7}},
  \bibinfo{pages}{10694} (\bibinfo{year}{2016}).

\bibitem{Mongera:2018aa}
\bibinfo{author}{Mongera, A.} \emph{et~al.}
\newblock \bibinfo{title}{A fluid-to-solid jamming transition underlies
  vertebrate body axis elongation}.
\newblock \emph{\bibinfo{journal}{Nature}} \textbf{\bibinfo{volume}{561}},
  \bibinfo{pages}{401--405} (\bibinfo{year}{2018}).

\bibitem{Bi:2015aa}
\bibinfo{author}{Bi, D.}, \bibinfo{author}{Lopez, J.~H.},
  \bibinfo{author}{Schwarz, J.~M.} \& \bibinfo{author}{Manning, M.~L.}
\newblock \bibinfo{title}{A density-independent rigidity transition in
  biological tissues}.
\newblock \emph{\bibinfo{journal}{Nat. Phys.}} \textbf{\bibinfo{volume}{11}},
  \bibinfo{pages}{1074} (\bibinfo{year}{2015}).

\bibitem{volfson2008biomechanical}
\bibinfo{author}{Volfson, D.}, \bibinfo{author}{Cookson, S.},
  \bibinfo{author}{Hasty, J.} \& \bibinfo{author}{Tsimring, L.~S.}
\newblock \bibinfo{title}{Biomechanical ordering of dense cell populations}.
\newblock \emph{\bibinfo{journal}{Proc. Natl. Acad. Sci. U.S.A.}}
  \textbf{\bibinfo{volume}{105}}, \bibinfo{pages}{15346--15351}
  (\bibinfo{year}{2008}).

\bibitem{ramananarivo2019activity}
\bibinfo{author}{Ramananarivo, S.}, \bibinfo{author}{Ducrot, E.} \&
  \bibinfo{author}{Palacci, J.}
\newblock \bibinfo{title}{Activity-controlled annealing of colloidal
  monolayers}.
\newblock \emph{\bibinfo{journal}{Nat. Commun.}} \textbf{\bibinfo{volume}{10}},
  \bibinfo{pages}{1--8} (\bibinfo{year}{2019}).

\bibitem{praetorius2018active}
\bibinfo{author}{Praetorius, S.}, \bibinfo{author}{Voigt, A.},
  \bibinfo{author}{Wittkowski, R.} \& \bibinfo{author}{L{\"o}wen, H.}
\newblock \bibinfo{title}{Active crystals on a sphere}.
\newblock \emph{\bibinfo{journal}{Phys. Rev. E}} \textbf{\bibinfo{volume}{97}},
  \bibinfo{pages}{052615} (\bibinfo{year}{2018}).

\bibitem{singh2016universal}
\bibinfo{author}{Singh, R.} \& \bibinfo{author}{Adhikari, R.}
\newblock \bibinfo{title}{Universal hydrodynamic mechanisms for crystallization
  in active colloidal suspensions}.
\newblock \emph{\bibinfo{journal}{Phys. Rev. Lett.}}
  \textbf{\bibinfo{volume}{117}}, \bibinfo{pages}{228002}
  (\bibinfo{year}{2016}).

\bibitem{van2016spatiotemporal}
\bibinfo{author}{van Zuiden, B.~C.}, \bibinfo{author}{Paulose, J.},
  \bibinfo{author}{Irvine, W.~T.}, \bibinfo{author}{Bartolo, D.} \&
  \bibinfo{author}{Vitelli, V.}
\newblock \bibinfo{title}{Spatiotemporal order and emergent edge currents in
  active spinner materials}.
\newblock \emph{\bibinfo{journal}{Proc. Natl. Acad. Sci. U.S.A.}}
  \textbf{\bibinfo{volume}{113}}, \bibinfo{pages}{12919--12924}
  (\bibinfo{year}{2016}).

\bibitem{nguyen2014emergent}
\bibinfo{author}{Nguyen, N.~H.}, \bibinfo{author}{Klotsa, D.},
  \bibinfo{author}{Engel, M.} \& \bibinfo{author}{Glotzer, S.~C.}
\newblock \bibinfo{title}{Emergent collective phenomena in a mixture of hard
  shapes through active rotation}.
\newblock \emph{\bibinfo{journal}{Phys. Rev. Lett.}}
  \textbf{\bibinfo{volume}{112}}, \bibinfo{pages}{075701}
  (\bibinfo{year}{2014}).

\bibitem{engel2013hard}
\bibinfo{author}{Engel, M.} \emph{et~al.}
\newblock \bibinfo{title}{Hard-disk equation of state: First-order
  liquid-hexatic transition in two dimensions with three simulation methods}.
\newblock \emph{\bibinfo{journal}{Phys. Rev. E}} \textbf{\bibinfo{volume}{87}},
  \bibinfo{pages}{042134} (\bibinfo{year}{2013}).

\bibitem{menzel2014active}
\bibinfo{author}{Menzel, A.~M.}, \bibinfo{author}{Ohta, T.} \&
  \bibinfo{author}{L{\"o}wen, H.}
\newblock \bibinfo{title}{Active crystals and their stability}.
\newblock \emph{\bibinfo{journal}{Phys. Rev. E}} \textbf{\bibinfo{volume}{89}},
  \bibinfo{pages}{022301} (\bibinfo{year}{2014}).

\bibitem{menzel2013traveling}
\bibinfo{author}{Menzel, A.~M.} \& \bibinfo{author}{L{\"o}wen, H.}
\newblock \bibinfo{title}{Traveling and resting crystals in active systems}.
\newblock \emph{\bibinfo{journal}{Phys. Rev. Lett.}}
  \textbf{\bibinfo{volume}{110}}, \bibinfo{pages}{055702}
  (\bibinfo{year}{2013}).

\bibitem{bialke2012crystallization}
\bibinfo{author}{Bialk{\'e}, J.}, \bibinfo{author}{Speck, T.} \&
  \bibinfo{author}{L{\"o}wen, H.}
\newblock \bibinfo{title}{Crystallization in a dense suspension of
  self-propelled particles}.
\newblock \emph{\bibinfo{journal}{Phys. Rev. Lett.}}
  \textbf{\bibinfo{volume}{108}}, \bibinfo{pages}{168301}
  (\bibinfo{year}{2012}).

\bibitem{durand2019thermally}
\bibinfo{author}{Durand, M.} \& \bibinfo{author}{Heu, J.}
\newblock \bibinfo{title}{Thermally driven order-disorder transition in
  two-dimensional soft cellular systems}.
\newblock \emph{\bibinfo{journal}{Phys. Rev. Lett.}}
  \textbf{\bibinfo{volume}{123}}, \bibinfo{pages}{188001}
  (\bibinfo{year}{2019}).

\bibitem{loewe2019solid}
\bibinfo{author}{Loewe, B.}, \bibinfo{author}{Chiang, M.},
  \bibinfo{author}{Marenduzzo, D.} \& \bibinfo{author}{Marchetti, M.~C.}
\newblock \bibinfo{title}{Solid-liquid transition of deformable and overlapping
  active particles}.
\newblock \emph{\bibinfo{journal}{arXiv preprint arXiv:1912.10549}}
  (\bibinfo{year}{2019}).

\bibitem{mermin1966absence}
\bibinfo{author}{Mermin, N.~D.} \& \bibinfo{author}{Wagner, H.}
\newblock \bibinfo{title}{Absence of ferromagnetism or antiferromagnetism in
  one-or two-dimensional isotropic {H}eisenberg models}.
\newblock \emph{\bibinfo{journal}{Phys. Rev. Lett.}}
  \textbf{\bibinfo{volume}{17}}, \bibinfo{pages}{1133} (\bibinfo{year}{1966}).

\bibitem{hohenberg1967existence}
\bibinfo{author}{Hohenberg, P.~C.}
\newblock \bibinfo{title}{Existence of long-range order in one and two
  dimensions}.
\newblock \emph{\bibinfo{journal}{Phys. Rev.}} \textbf{\bibinfo{volume}{158}},
  \bibinfo{pages}{383} (\bibinfo{year}{1967}).

\bibitem{mermin1968crystalline}
\bibinfo{author}{Mermin, N.~D.}
\newblock \bibinfo{title}{Crystalline order in two dimensions}.
\newblock \emph{\bibinfo{journal}{Phys. Rev.}} \textbf{\bibinfo{volume}{176}},
  \bibinfo{pages}{250} (\bibinfo{year}{1968}).

\bibitem{halperin2019jstat}
\bibinfo{author}{Halperin, B.~I.}
\newblock \bibinfo{title}{On the {H}ohenberg--{M}ermin--{W}agner theorem and
  its limitations}.
\newblock \emph{\bibinfo{journal}{J. Stat. Phys.}}
  \textbf{\bibinfo{volume}{175}}, \bibinfo{pages}{521--529}
  (\bibinfo{year}{2019}).

\bibitem{ramakrishnan1979first}
\bibinfo{author}{Ramakrishnan, T.} \& \bibinfo{author}{Yussouff, M.}
\newblock \bibinfo{title}{First-principles order-parameter theory of freezing}.
\newblock \emph{\bibinfo{journal}{Phys. Rev. B}} \textbf{\bibinfo{volume}{19}},
  \bibinfo{pages}{2775} (\bibinfo{year}{1979}).

\bibitem{chui1983grain}
\bibinfo{author}{Chui, S.}
\newblock \bibinfo{title}{Grain-boundary theory of melting in two dimensions}.
\newblock \emph{\bibinfo{journal}{Phys. Rev. B}} \textbf{\bibinfo{volume}{28}},
  \bibinfo{pages}{178} (\bibinfo{year}{1983}).

\bibitem{halperin1978theory}
\bibinfo{author}{Halperin, B.} \& \bibinfo{author}{Nelson, D.~R.}
\newblock \bibinfo{title}{Theory of two-dimensional melting}.
\newblock \emph{\bibinfo{journal}{Phys. Rev. Lett.}}
  \textbf{\bibinfo{volume}{41}}, \bibinfo{pages}{121} (\bibinfo{year}{1978}).

\bibitem{zahn2000dynamic}
\bibinfo{author}{Zahn, K.} \& \bibinfo{author}{Maret, G.}
\newblock \bibinfo{title}{Dynamic criteria for melting in two dimensions}.
\newblock \emph{\bibinfo{journal}{Phys. Rev. Lett.}}
  \textbf{\bibinfo{volume}{85}}, \bibinfo{pages}{3656} (\bibinfo{year}{2000}).

\bibitem{guillamon2009direct}
\bibinfo{author}{Guillam{\'o}n, I.} \emph{et~al.}
\newblock \bibinfo{title}{Direct observation of melting in a two-dimensional
  superconducting vortex lattice}.
\newblock \emph{\bibinfo{journal}{Nat. Phys.}} \textbf{\bibinfo{volume}{5}},
  \bibinfo{pages}{651} (\bibinfo{year}{2009}).

\bibitem{kapfer2015two}
\bibinfo{author}{Kapfer, S.~C.} \& \bibinfo{author}{Krauth, W.}
\newblock \bibinfo{title}{Two-dimensional melting: From liquid-hexatic
  coexistence to continuous transitions}.
\newblock \emph{\bibinfo{journal}{Phys. Rev. Lett.}}
  \textbf{\bibinfo{volume}{114}}, \bibinfo{pages}{035702}
  (\bibinfo{year}{2015}).

\bibitem{goldman2003lattice}
\bibinfo{author}{Goldman, D.~I.}, \bibinfo{author}{Shattuck, M.},
  \bibinfo{author}{Moon, S.~J.}, \bibinfo{author}{Swift, J.} \&
  \bibinfo{author}{Swinney, H.~L.}
\newblock \bibinfo{title}{Lattice dynamics and melting of a nonequilibrium
  pattern}.
\newblock \emph{\bibinfo{journal}{Phys. Rev. Lett.}}
  \textbf{\bibinfo{volume}{90}}, \bibinfo{pages}{104302}
  (\bibinfo{year}{2003}).

\bibitem{boyer2009two}
\bibinfo{author}{Boyer, F.} \& \bibinfo{author}{Falcon, E.}
\newblock \bibinfo{title}{Two-dimensional melting of a crystal of ferrofluid
  spikes}.
\newblock \emph{\bibinfo{journal}{Phys. Rev. Lett.}}
  \textbf{\bibinfo{volume}{103}}, \bibinfo{pages}{144501}
  (\bibinfo{year}{2009}).

\bibitem{perlekar2010turbulence}
\bibinfo{author}{Perlekar, P.} \& \bibinfo{author}{Pandit, R.}
\newblock \bibinfo{title}{Turbulence-induced melting of a nonequilibrium vortex
  crystal in a forced thin fluid film}.
\newblock \emph{\bibinfo{journal}{New J. Phys.}} \textbf{\bibinfo{volume}{12}},
  \bibinfo{pages}{023033} (\bibinfo{year}{2010}).

\bibitem{wensink2012meso}
\bibinfo{author}{Wensink, H.~H.} \emph{et~al.}
\newblock \bibinfo{title}{Meso-scale turbulence in living fluids}.
\newblock \emph{\bibinfo{journal}{Proc. Natl. Acad. Sci. U.S.A.}}
  \textbf{\bibinfo{volume}{109}}, \bibinfo{pages}{14308--14313}
  (\bibinfo{year}{2012}).

\bibitem{dunkel2013minimal}
\bibinfo{author}{Dunkel, J.}, \bibinfo{author}{Heidenreich, S.},
  \bibinfo{author}{B{\"a}r, M.} \& \bibinfo{author}{Goldstein, R.~E.}
\newblock \bibinfo{title}{Minimal continuum theories of structure formation in
  dense active fluids}.
\newblock \emph{\bibinfo{journal}{New J. Phys.}} \textbf{\bibinfo{volume}{15}},
  \bibinfo{pages}{045016} (\bibinfo{year}{2013}).

\bibitem{doostmohammadi2017onset}
\bibinfo{author}{Doostmohammadi, A.}, \bibinfo{author}{Shendruk, T.~N.},
  \bibinfo{author}{Thijssen, K.} \& \bibinfo{author}{Yeomans, J.~M.}
\newblock \bibinfo{title}{Onset of meso-scale turbulence in active nematics}.
\newblock \emph{\bibinfo{journal}{Nat. Commun.}} \textbf{\bibinfo{volume}{8}},
  \bibinfo{pages}{15326} (\bibinfo{year}{2017}).

\bibitem{james2018vortex}
\bibinfo{author}{James, M.} \& \bibinfo{author}{Wilczek, M.}
\newblock \bibinfo{title}{Vortex dynamics and {L}agrangian statistics in a
  model for active turbulence}.
\newblock \emph{\bibinfo{journal}{Eur. Phys. J. E}}
  \textbf{\bibinfo{volume}{41}}, \bibinfo{pages}{21} (\bibinfo{year}{2018}).

\bibitem{james2018turbulence}
\bibinfo{author}{James, M.}, \bibinfo{author}{Bos, W.~J.} \&
  \bibinfo{author}{Wilczek, M.}
\newblock \bibinfo{title}{Turbulence and turbulent pattern formation in a
  minimal model for active fluids}.
\newblock \emph{\bibinfo{journal}{Phys. Rev. Fluids}}
  \textbf{\bibinfo{volume}{3}}, \bibinfo{pages}{061101} (\bibinfo{year}{2018}).

\bibitem{sumino2012large}
\bibinfo{author}{Sumino, Y.} \emph{et~al.}
\newblock \bibinfo{title}{Large-scale vortex lattice emerging from collectively
  moving microtubules}.
\newblock \emph{\bibinfo{journal}{Nature}} \textbf{\bibinfo{volume}{483}},
  \bibinfo{pages}{448} (\bibinfo{year}{2012}).

\bibitem{shendruk2017dancing}
\bibinfo{author}{Shendruk, T.~N.}, \bibinfo{author}{Doostmohammadi, A.},
  \bibinfo{author}{Thijssen, K.} \& \bibinfo{author}{Yeomans, J.~M.}
\newblock \bibinfo{title}{Dancing disclinations in confined active nematics}.
\newblock \emph{\bibinfo{journal}{Soft Matter}} \textbf{\bibinfo{volume}{13}},
  \bibinfo{pages}{3853--3862} (\bibinfo{year}{2017}).

\bibitem{2017SlomkaDunkel_PRF}
\bibinfo{author}{S\l{}omka, J.} \& \bibinfo{author}{Dunkel, J.}
\newblock \bibinfo{title}{Geometry-dependent viscosity reduction in sheared
  active fluids}.
\newblock \emph{\bibinfo{journal}{Phys. Rev. Fluids}}
  \textbf{\bibinfo{volume}{2}}, \bibinfo{pages}{043102} (\bibinfo{year}{2017}).

\bibitem{Doostmohammadi:2016aa}
\bibinfo{author}{Doostmohammadi, A.}, \bibinfo{author}{Adamer, M.~F.},
  \bibinfo{author}{Thampi, S.~P.} \& \bibinfo{author}{Yeomans, J.~M.}
\newblock \bibinfo{title}{Stabilization of active matter by flow-vortex
  lattices and defect ordering}.
\newblock \emph{\bibinfo{journal}{Nat. Commun.}} \textbf{\bibinfo{volume}{7}},
  \bibinfo{pages}{10557} (\bibinfo{year}{2016}).

\bibitem{2016OzHeDu_EPJE}
\bibinfo{author}{Oza, A.~U.}, \bibinfo{author}{Heidenreich, S.} \&
  \bibinfo{author}{Dunkel, J.}
\newblock \bibinfo{title}{Generalized {Swift-Hohenberg} equations for dense
  active suspensions}.
\newblock \emph{\bibinfo{journal}{Eur. Phys. J. E}}
  \textbf{\bibinfo{volume}{39}}, \bibinfo{pages}{97} (\bibinfo{year}{2016}).

\bibitem{nagai2015collective}
\bibinfo{author}{Nagai, K.~H.}, \bibinfo{author}{Sumino, Y.},
  \bibinfo{author}{Montagne, R.}, \bibinfo{author}{Aranson, I.~S.} \&
  \bibinfo{author}{Chat{\'e}, H.}
\newblock \bibinfo{title}{Collective motion of self-propelled particles with
  memory}.
\newblock \emph{\bibinfo{journal}{Phys. Rev. Lett.}}
  \textbf{\bibinfo{volume}{114}}, \bibinfo{pages}{168001}
  (\bibinfo{year}{2015}).

\bibitem{grossmann2014vortex}
\bibinfo{author}{Gro{\ss}mann, R.}, \bibinfo{author}{Romanczuk, P.},
  \bibinfo{author}{B{\"a}r, M.} \& \bibinfo{author}{Schimansky-Geier, L.}
\newblock \bibinfo{title}{Vortex arrays and mesoscale turbulence of
  self-propelled particles}.
\newblock \emph{\bibinfo{journal}{Phys. Rev. Lett.}}
  \textbf{\bibinfo{volume}{113}}, \bibinfo{pages}{258104}
  (\bibinfo{year}{2014}).

\bibitem{bratanov2015new}
\bibinfo{author}{Bratanov, V.}, \bibinfo{author}{Jenko, F.} \&
  \bibinfo{author}{Frey, E.}
\newblock \bibinfo{title}{New class of turbulence in active fluids}.
\newblock \emph{\bibinfo{journal}{Proc. Natl. Acad. Sci. U.S.A.}}
  \textbf{\bibinfo{volume}{112}}, \bibinfo{pages}{15048--15053}
  (\bibinfo{year}{2015}).

\bibitem{toner1998flocks}
\bibinfo{author}{Toner, J.} \& \bibinfo{author}{Tu, Y.}
\newblock \bibinfo{title}{Flocks, herds, and schools: A quantitative theory of
  flocking}.
\newblock \emph{\bibinfo{journal}{Phys. Rev. E}} \textbf{\bibinfo{volume}{58}},
  \bibinfo{pages}{4828} (\bibinfo{year}{1998}).

\bibitem{toner2005hydrodynamics}
\bibinfo{author}{Toner, J.}, \bibinfo{author}{Tu, Y.} \&
  \bibinfo{author}{Ramaswamy, S.}
\newblock \bibinfo{title}{Hydrodynamics and phases of flocks}.
\newblock \emph{\bibinfo{journal}{Ann. Phys.}} \textbf{\bibinfo{volume}{318}},
  \bibinfo{pages}{170--244} (\bibinfo{year}{2005}).

\bibitem{marchetti2013hydrodynamics}
\bibinfo{author}{Marchetti, M.~C.} \emph{et~al.}
\newblock \bibinfo{title}{Hydrodynamics of soft active matter}.
\newblock \emph{\bibinfo{journal}{Rev. Mod. Phys.}}
  \textbf{\bibinfo{volume}{85}}, \bibinfo{pages}{1143} (\bibinfo{year}{2013}).

\bibitem{2016HeEtAl_PRE}
\bibinfo{author}{Heidenreich, S.}, \bibinfo{author}{Dunkel, J.},
  \bibinfo{author}{Klapp, S. H.~L.} \& \bibinfo{author}{B\"ar, M.}
\newblock \bibinfo{title}{Hydrodynamic length-scale selection in microswimmer
  suspensions}.
\newblock \emph{\bibinfo{journal}{Phys. Rev. E}} \textbf{\bibinfo{volume}{94}},
  \bibinfo{pages}{020601(R)} (\bibinfo{year}{2016}).

\bibitem{PhysRevE.97.022613}
\bibinfo{author}{Reinken, H.}, \bibinfo{author}{Klapp, S. H.~L.},
  \bibinfo{author}{B{\"a}r, M.} \& \bibinfo{author}{Heidenreich, S.}
\newblock \bibinfo{title}{Derivation of a hydrodynamic theory for mesoscale
  dynamics in microswimmer suspensions}.
\newblock \emph{\bibinfo{journal}{Phys. Rev. E}} \textbf{\bibinfo{volume}{97}},
  \bibinfo{pages}{022613} (\bibinfo{year}{2018}).

\bibitem{simha2002hydrodynamic}
\bibinfo{author}{Simha, R.~A.} \& \bibinfo{author}{Ramaswamy, S.}
\newblock \bibinfo{title}{Hydrodynamic fluctuations and instabilities in
  ordered suspensions of self-propelled particles}.
\newblock \emph{\bibinfo{journal}{Phys. Rev. Lett.}}
  \textbf{\bibinfo{volume}{89}}, \bibinfo{pages}{058101}
  (\bibinfo{year}{2002}).

\bibitem{saintillan2008instabilities}
\bibinfo{author}{Saintillan, D.} \& \bibinfo{author}{Shelley, M.~J.}
\newblock \bibinfo{title}{Instabilities and pattern formation in active
  particle suspensions: kinetic theory and continuum simulations}.
\newblock \emph{\bibinfo{journal}{Phys. Rev. Lett.}}
  \textbf{\bibinfo{volume}{100}}, \bibinfo{pages}{178103}
  (\bibinfo{year}{2008}).

\bibitem{bedanov1985modified}
\bibinfo{author}{Bedanov, V.}, \bibinfo{author}{Gadiyak, G.} \&
  \bibinfo{author}{Lozovik, Y.~E.}
\newblock \bibinfo{title}{On a modified {L}indemann-like criterion for 2d
  melting}.
\newblock \emph{\bibinfo{journal}{Phys. Lett. A}}
  \textbf{\bibinfo{volume}{109}}, \bibinfo{pages}{289--291}
  (\bibinfo{year}{1985}).

\bibitem{zahn1999two}
\bibinfo{author}{Zahn, K.}, \bibinfo{author}{Lenke, R.} \&
  \bibinfo{author}{Maret, G.}
\newblock \bibinfo{title}{Two-stage melting of paramagnetic colloidal crystals
  in two dimensions}.
\newblock \emph{\bibinfo{journal}{Phys. Rev. Lett.}}
  \textbf{\bibinfo{volume}{82}}, \bibinfo{pages}{2721} (\bibinfo{year}{1999}).

\bibitem{kaneko1990supertransients}
\bibinfo{author}{Kaneko, K.}
\newblock \bibinfo{title}{Supertransients, spatiotemporal intermittency and
  stability of fully developed spatiotemporal chaos}.
\newblock \emph{\bibinfo{journal}{Phys. Lett. A}}
  \textbf{\bibinfo{volume}{149}}, \bibinfo{pages}{105--112}
  (\bibinfo{year}{1990}).

\bibitem{crutchfield1988attractors}
\bibinfo{author}{Crutchfield, J.~P.} \& \bibinfo{author}{Kaneko, K.}
\newblock \bibinfo{title}{Are attractors relevant to turbulence?}
\newblock \emph{\bibinfo{journal}{Phys. Rev. Lett.}}
  \textbf{\bibinfo{volume}{60}}, \bibinfo{pages}{2715} (\bibinfo{year}{1988}).

\bibitem{strain1998size}
\bibinfo{author}{Strain, M.~C.} \& \bibinfo{author}{Greenside, H.~S.}
\newblock \bibinfo{title}{Size-dependent transition to high-dimensional chaotic
  dynamics in a two-dimensional excitable medium}.
\newblock \emph{\bibinfo{journal}{Phys. Rev. Lett.}}
  \textbf{\bibinfo{volume}{80}}, \bibinfo{pages}{2306} (\bibinfo{year}{1998}).

\bibitem{lai2011transient}
\bibinfo{author}{Lai, Y.-C.} \& \bibinfo{author}{T{\'e}l, T.}
\newblock \emph{\bibinfo{title}{Transient chaos: complex dynamics on finite
  time scales}} (\bibinfo{publisher}{Springer Science \& Business Media,
  Springer, New York, NY}, \bibinfo{year}{2011}).

\bibitem{creppy2015turbulence}
\bibinfo{author}{Creppy, A.}, \bibinfo{author}{Praud, O.},
  \bibinfo{author}{Druart, X.}, \bibinfo{author}{Kohnke, P.~L.} \&
  \bibinfo{author}{Plourabou{\'e}, F.}
\newblock \bibinfo{title}{Turbulence of swarming sperm}.
\newblock \emph{\bibinfo{journal}{Phys. Rev. E}} \textbf{\bibinfo{volume}{92}},
  \bibinfo{pages}{032722} (\bibinfo{year}{2015}).

\bibitem{fisher2014dynamics}
\bibinfo{author}{Fisher, H.~S.}, \bibinfo{author}{Giomi, L.},
  \bibinfo{author}{Hoekstra, H.~E.} \& \bibinfo{author}{Mahadevan, L.}
\newblock \bibinfo{title}{The dynamics of sperm cooperation in a competitive
  environment}.
\newblock \emph{\bibinfo{journal}{Proc. R. Soc. B}}
  \textbf{\bibinfo{volume}{281}}, \bibinfo{pages}{20140296}
  (\bibinfo{year}{2014}).

\bibitem{alvarez2012rate}
\bibinfo{author}{Alvarez, L.} \emph{et~al.}
\newblock \bibinfo{title}{The rate of change in {Ca2+} concentration controls
  sperm chemotaxis}.
\newblock \emph{\bibinfo{journal}{J. Cell Biol.}}
  \textbf{\bibinfo{volume}{196}}, \bibinfo{pages}{653--663}
  (\bibinfo{year}{2012}).

\bibitem{alavi2005sperm}
\bibinfo{author}{Alavi, S. M.~H.} \& \bibinfo{author}{Cosson, J.}
\newblock \bibinfo{title}{Sperm motility in fishes. {I. Effects of temperature
  and pH: a review}}.
\newblock \emph{\bibinfo{journal}{Cell Biol. Int.}}
  \textbf{\bibinfo{volume}{29}}, \bibinfo{pages}{101--110}
  (\bibinfo{year}{2005}).

\bibitem{woolley2003motility}
\bibinfo{author}{Woolley, D.}
\newblock \bibinfo{title}{Motility of spermatozoa at surfaces}.
\newblock \emph{\bibinfo{journal}{Reproduction}}
  \textbf{\bibinfo{volume}{126}}, \bibinfo{pages}{259--270}
  (\bibinfo{year}{2003}).

\bibitem{giomi2015geometry}
\bibinfo{author}{Giomi, L.}
\newblock \bibinfo{title}{Geometry and topology of turbulence in active
  nematics}.
\newblock \emph{\bibinfo{journal}{Phys. Rev. X}} \textbf{\bibinfo{volume}{5}},
  \bibinfo{pages}{031003} (\bibinfo{year}{2015}).

\end{thebibliography}
\newpage
\begin{acknowledgments}

This work was supported by the Max Planck Society. M.W. gratefully acknowledges a Fulbright-Cottrell Award grant. M.J. gratefully acknowledges financial support through an IMPRS-PBCS fellowship. M.J. thanks Stephan Herminghaus and Marcus M\"{u}ller for helpful discussions.
\end{acknowledgments}

\section*{Author Contributions}

M.J. and M.W. designed the research. D.A.S. wrote the DNS code. M.J. and D.A.S. conducted the analysis. All authors contributed to the interpretation of the results and the writing of the manuscript
\newpage
\section*{Additional Information}
\subsection*{List of supplementary videos.}

\begin{enumerate}[label=\textnormal{(\arabic*)}]
    \item \href{https://youtu.be/p_br0pkBEVs}{Evolution of AVC superstructures. The inset shows a zoom-in into the turbulent interfacial area.}\label{vid:superstructureevolution}
    \item \href{https://youtu.be/n-2Yue-WFWc}{Emergence of a vortex crystal starting from random initial conditions.}\label{vid:vortexarrayemergence}
    \item \href{https://youtu.be/jQZlmNoSzFs}{The active matter system in the marginal stability region between the turbulent phase and the vortex crystal. The right panel shows how the energy density changes with time.}\label{vid:marginalstability}
\end{enumerate}

\subsection*{Competing financial interests.} The authors declare no competing financial interests.

\end{document}